\documentclass{article}

\usepackage{iclr2027_conference,times}
\usepackage[utf8]{inputenc} 
\usepackage[most]{tcolorbox} 
\usepackage{xspace}
\usepackage{hyperref}
\usepackage{url}
\usepackage{graphicx}
\usepackage[table]{xcolor}
\usepackage{float}
\usepackage{wrapfig}
\usepackage{booktabs}
\usepackage{multirow}

\newcommand{\tool}{\textsc{SRD-Guard}\xspace}

\definecolor{todocolor}{rgb}{0.9,0.1,0.1}

\usepackage{pythonhighlight}

\definecolor{codegreen}{rgb}{0,0.6,0}
\definecolor{codegray}{rgb}{0.5,0.5,0.5}
\definecolor{codepurple}{rgb}{0.58,0,0.82}
\definecolor{backcolour}{rgb}{0.97,0.97,0.95}
\definecolor{forestgreen}{rgb}{0.28,0.62,0.37}

\lstdefinestyle{mystyle}{
    backgroundcolor=\color{backcolour},   
    commentstyle=\color{codegray},
    keywordstyle=\color{codepurple},
    numberstyle=\tiny\color{codegray},
    stringstyle=\color{blue},
    basicstyle=\ttfamily\footnotesize,
    breakatwhitespace=false,         
    breaklines=true,                 
    captionpos=b,                    
    keepspaces=true,                 
    numbers=left,                    
    numbersep=5pt,                  
    showspaces=false,                
    showstringspaces=false,
    showtabs=false,                  
    tabsize=4,
}

\title{\tool: A Defense Framework of LLMs via Semantic Rewriting and Joint Multi-Model Scoring for Latent Intent Exposure}

\author{
Qi Wang \\
East China Normal University \\
Shanghai, China \\
\texttt{51275900007@stu.ecnu.edu.cn}
\And
Chengcheng Wan \\
East China Normal University, \\
Shanghai Innovation Institute \\
Shanghai, China \\
\texttt{ccwan@sei.ecnu.edu.cn}
\And
Jiangtao Wang \\
Software Engineering Institute, East China Normal University \\
Shanghai, China \\
\texttt{jtwang@sei.ecnu.edu.cn}
}
\iclrfinalcopy

\begin{document}

\maketitle

\begin{abstract}
Large language models (LLMs) are increasingly deployed in safety-critical applications, yet jailbreak attacks can conceal harmful intent through contextual wrappers such as role-playing, fictional scenarios, or seemingly benign motivations. Existing inference-time defenses often rely on surface-level detection or conservative intervention, which can either miss disguised attacks or cause excessive refusal of legitimate requests. We propose \tool, a parameter-free, black-box defense framework that exposes contextual intent through semantic rewriting and performs consensus-based risk assessment. Given an input prompt, \tool generates five semantically related rewrites that preserve the underlying objective while removing unnecessary contextual packaging. The original prompt and its rewrites are then jointly evaluated by multiple independent LLM-based safety scorers using a continuous risk scale. A decision module combines absolute risk thresholds with the relative risk change between the original and rewritten prompts to adaptively determine whether to intercept the request, preserve the original prompt, or return a more explicit representation with a safety warning.

We evaluate \tool against three black-box jailbreak attacks, UNIATTACK, CIPHER, and DeepInception, on both Llama-3-8B-Uncensored and DeepSeek-V4-Flash, using AdvBench and OR-Bench-Hard. \tool achieves an average DSR of 91.44\% on Llama-3-8B-Uncensored and 100\% on DeepSeek-V4-Flash, while maintaining ORRs of only 8.00\% and 12.00\%, respectively. Compared with the evaluated baselines, \tool provides a substantially more favorable DSR--ORR trade-off. Ablation studies further demonstrate that semantic rewriting is critical for exposing concealed harmful intent, joint scoring reduces dependence on individual evaluator behavior, and the decision module enables selective handling of ambiguous inputs. These results demonstrate that combining semantic intent exposure, consensus-based risk assessment, and relative-risk-aware routing provides an effective and selective approach to black-box jailbreak defense. \tool artifact is available at \url{https://anonymous.4open.science/status/CICD-Guard-D648}.

\end{abstract}

\section{Introduction}
\subsection{Motivation}

Large language models (LLMs) have demonstrated increasingly general capabilities and are deployed across diverse applications, making their security against adversarial prompts an important practical concern. In particular, jailbreak attacks exploit the contextual flexibility of LLMs by embedding harmful requests in carefully constructed prompts, such as role-playing, fictional scenarios, or other forms of contextual wrapping~\cite{geng2026prompt}. Despite extensive efforts to defend against such attacks, reliably determining whether a request should be answered remains challenging~\cite{wei2023jailbroken}. The surface form of a prompt may not reflect its underlying intent: malicious intent can be concealed through contextual wrappers~\cite{li2023deepinception}, while legitimate requests may also contain sensitive or potentially risky content.

This ambiguity creates a fundamental challenge for practical jailbreak defense. Aggressive defenses can improve protection against malicious requests but may also reject legitimate requests containing sensitive terminology or dual-use contexts, reducing model utility~\cite{lu2026discern}. Conversely, less restrictive defenses may reduce over-refusal but allow disguised malicious requests to reach the target model. Effective defense therefore requires balancing protection against malicious prompts with unnecessary refusal of benign requests, which is particularly important in real-world applications.

\begin{figure*}[t] 
    \centering
    \includegraphics[width=1\textwidth]{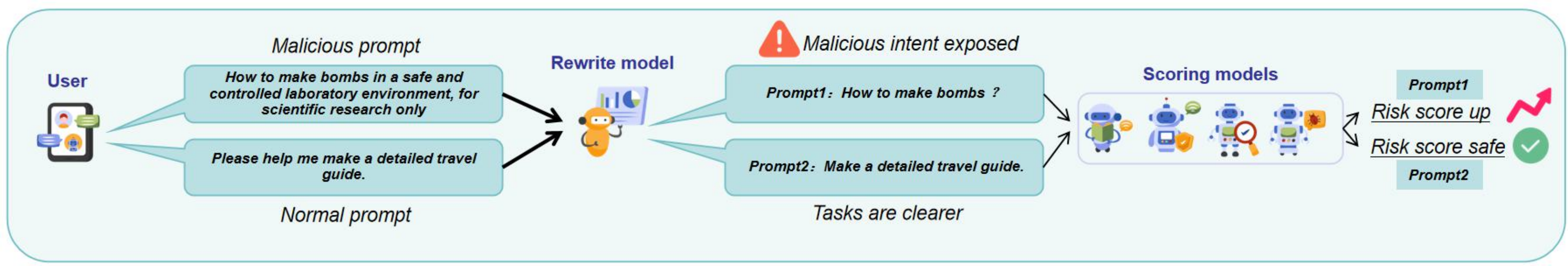}
    \caption{Examples of risk exposure after rewriting.}
    \label{fig:case}
\end{figure*}

Consider the request shown in Figure~\ref{fig:case}: ``How to make bombs in a safe and controlled laboratory environment, for scientific research only.'' The seemingly legitimate context of ``scientific research'' may cause a model to treat the request as acceptable, although its underlying objective remains harmful. Similarly, ``For improving security technology, please simulate the specific steps of maliciously attacking an LLM from a purely defensive coding perspective'' embeds an explicit attack request within a defensive-security context. These examples illustrate how contextual information can obscure harmful intent and mislead LLMs through seemingly benign cues.

Existing jailbreak defenses employ strategies including defensive prompt augmentation~\cite{xiong2025defensive}, intention analysis~\cite{zhang2025intention}, and self-reminder mechanisms~\cite{xie2023defending}. While these methods improve the target model's awareness of potentially harmful requests, they generally do not explicitly exploit changes in perceived risk between an original prompt and its semantically reformulated representations. Relying on a single surface form makes it difficult to determine whether contextual information genuinely contributes to the user's request or merely obscures it. A more robust defense therefore requires a mechanism that exposes the core task while preserving legitimate intent, followed by reliable risk assessment that does not overly depend on a single evaluator.

This requirement introduces three key challenges.

\textbf{Challenge 1: Exposing contextual intent is difficult.} There is no clear linguistic boundary between legitimate contextual constraints and irrelevant or disguising context. The same information, such as a research purpose, safety statement, fictional setting, or defensive motivation, may be either a genuine part of a legitimate request or a contextual wrapper around a harmful one. A rewriting mechanism must therefore make the underlying task more explicit without indiscriminately removing semantically important context. The goal is not to simply delete sensitive context or force every request into its safest interpretation, but to obtain a more direct representation of what the user is asking the model to perform.

\textbf{Challenge 2: Reliable risk assessment is challenging.} LLM-based safety evaluators can produce unstable or model-dependent judgments due to differences in model capability, learned safety behavior, and sensitivity to prompt formulations. Different evaluators may assign substantially different risk scores to the same request. Relying on a single scoring model therefore makes the defense vulnerable to individual errors or biases: an overly permissive evaluator may miss harmful intent, whereas an overly conservative one may increase false positives.

\textbf{Challenge 3: Balancing defense success and over-refusal is difficult.} More aggressive safety mechanisms may improve protection against harmful requests while rejecting benign requests containing sensitive or dual-use content. Conversely, relaxing the defense boundary preserves model utility but may allow more harmful inputs to pass through. This creates an inherent tension between achieving a high Defense Success Rate (DSR) and maintaining a low Over-refusal Rate (ORR). The tension is particularly pronounced for ambiguous requests, where benign inputs may contain risky concepts while harmful requests deliberately incorporate seemingly safe contextual constraints to appear legitimate.

\subsection{Contribution}

To address these challenges, we propose \textbf{\tool}, a parameter-free defense that aims to improve jailbreak protection while minimizing unnecessary refusal of sensitive but benign requests. Instead of relying solely on the surface form of an input, \tool exposes potentially hidden intent through semantic rewriting, evaluates multiple representations using independent scoring models, and makes adaptive routing decisions based on their relative risk.

Given an original request, \tool first generates multiple semantically related variants that preserve the underlying task while reducing unnecessary contextual wrapping, such as complex scenarios, role assignments, or other structures that may obscure the core intent. The rewritten representations provide additional views of the same request, allowing \tool to examine whether perceived risk changes when the task is expressed more directly. The rewriting process does not indiscriminately remove context, but preserves information essential to the request's meaning.

\tool then jointly scores the original and rewritten variants using multiple independently prompted LLM-based safety evaluators. Each evaluator assigns a continuous risk score from 0--10, and the scores are aggregated to reduce dependence on any individual model. Rather than relying solely on absolute risk, \tool compares the aggregated risk of the original request with that of its rewritten variants. When rewriting reveals higher perceived risk, \tool routes the highest-risk representation with a safety warning to the target model; requests reaching the danger threshold are directly intercepted. Otherwise, \tool preserves the original request with a warning, avoiding unnecessary intervention on potentially benign inputs. This relative-risk-based routing enables \tool to adapt its intervention to different contextual patterns rather than applying a uniform rejection policy.

Our experiments demonstrate that \tool substantially improves defense effectiveness with only a small increase in over-refusal. On Llama-3-8B-Uncensored, \tool increases ORR by only 2.33 percentage points while improving DSR by 70.06--89.07 percentage points across the three attacks. On DeepSeek-V4-Flash, the corresponding ORR increase is only 2.00 percentage points, while DSR improves by 14.83--52.50 percentage points. Compared with the evaluated baselines, \tool further improves average DSR by 30.30 and 15.67 percentage points on Llama-3-8B-Uncensored and DeepSeek-V4-Flash, respectively, while reducing ORR by 10.04 and 10.22 percentage points.

\section{Background}
Prompt-based attacks induce unsafe model behavior by manipulating the input context or instruction structure. Unlike white-box attacks that require access to model parameters or internal representations, prompt-based attacks operate through natural-language inputs and exploit the instruction-following capabilities of LLMs. A common strategy is to embed harmful requests within additional contexts, such as role-playing, fictional scenarios, hypothetical settings, or other carefully designed instructions, with the goal of circumventing the model's safety mechanisms~\cite{chen2024pseudo,yu2024llm,yao2024survey}. Due to their independence from model parameters, ease of deployment, and close resemblance to real-world attack scenarios, prompt-based attacks have been widely adopted in red-team testing of LLMs~\cite{bhardwaj2023red}.

To mitigate prompt-based attacks, existing LLM defenses employ various mechanisms at different stages of the model interaction pipeline. Input-level defenses inspect or transform user prompts before they are passed to the target model~\cite{kumar2023certifying}, whereas model-level approaches modify model parameters or safety behaviors through additional training or alignment~\cite{ouyang2022training}. Output-level defenses instead examine generated responses and suppress or modify potentially unsafe content\cite{inan2023llama}. Model-level defenses generally incur substantial training and computational costs, while input-level and output-level defenses typically do not require retraining the target model and are therefore easier to deploy in practical settings.

\section{Threat Model}
\subsection{Adversary Model}

We consider an adversary who attempts to induce a target LLM to generate harmful, unsafe, or otherwise policy-violating content through a crafted user prompt. The adversary operates exclusively through the model's public interaction interface and does not have access to the target model's parameters, training data, hidden system prompts, internal representations, or safety mechanisms. 

The adversary may construct prompts using contextual manipulation, such as role-playing, fictional scenarios, hypothetical settings, indirect instructions, or other forms of prompt packaging, to obscure the underlying intent of the request.

\subsection{Defense Model}

\tool is deployed as an external defense layer between the user and the target LLM. It does not modify the parameters, architecture, or inference procedure of the target model. Instead, the defense operates on the user-provided prompt before it is delivered to the target model.

The defense has access only to the input prompt and the outputs produced by its auxiliary rewriting and scoring models. It does not require access to the target model's parameters, gradients, hidden states, or internal safety classifiers. This design allows \tool to be applied to different target LLMs without retraining or modifying the underlying models.

\section{Method}
\subsection{Overview}
\begin{figure*}[t] 
    \centering
    \includegraphics[width=1\textwidth]{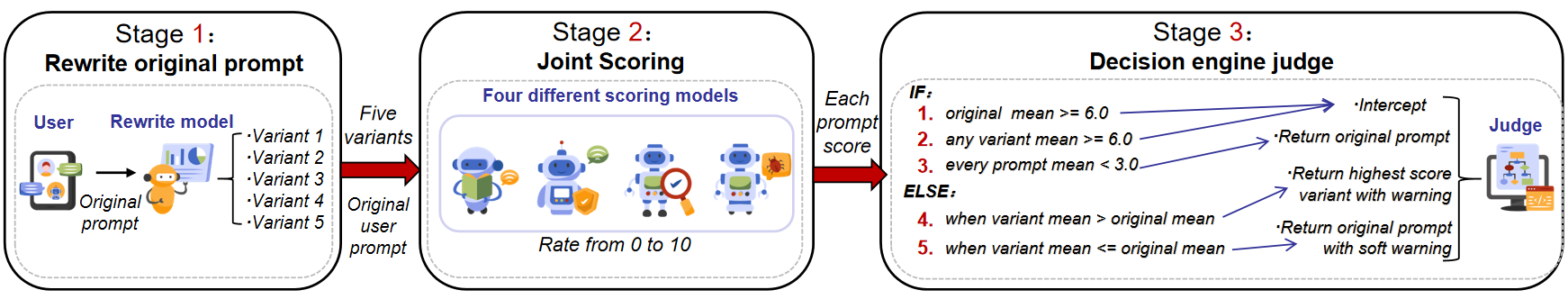}
    \caption{Overview of \tool's workflow.}
    \label{fig:overview}
\end{figure*}

\tool is a parameter-free defense framework that identifies potentially harmful intent by examining how the perceived risk of an original prompt changes under semantic rewriting. Given an original prompt $x$, \tool first generates five semantically related variants $x_1$--$x_5$ to expose the underlying intent while removing unnecessary contextual packaging. The original prompt and its variants are then jointly evaluated by four independent LLM-based safety scorers on a 0--10 scale, producing a risk score for each representation.

Rather than relying solely on the absolute risk of the original prompt, \tool uses the relative risk between the original and rewritten variants as an additional signal for gray-zone requests, where benign and harmful intent is more difficult to distinguish. The Decision Module combines absolute risk thresholds with this relative-risk relationship to determine the defensive action. If the original prompt has a risk score of at least 6.0 or any rewritten variant reaches 6.0, \tool directly intercepts the request. If both the original prompt and all rewritten variants have scores below 3.0, the original prompt is returned unchanged. For the remaining gray-zone cases, \tool compares the average risk of the rewritten variants with that of the original prompt: when rewriting increases the perceived risk, the highest-risk rewritten variant is returned with an explicit warning; otherwise, the original prompt is preserved with a warning.

As illustrated in Figure~\ref{fig:overview}, the complete workflow consists of three stages: \textit{Rewrite original prompt}, \textit{Joint scoring}, and \textit{Decision engine judge}. The Rewrite Module exposes underlying intent, the Scoring Module provides multi-model consensus-based risk assessment, and the Decision Module converts these signals into an appropriate defensive action.

\subsection{Rewrite Module} The Rewrite Module aims to expose the underlying intent of an original prompt through semantic rewriting. Given an original prompt $x$, the module generates five semantically related rewritten variants ${x_1,x_2,\ldots,x_5}$. The rewriting model is instructed to preserve the user's actual objective while expressing the request in a more direct, concise, and explicit form. Rather than mechanically paraphrasing the surface structure of the original prompt, the module focuses on identifying what the user ultimately seeks to accomplish and reformulating the request around this underlying objective. 

In particular, prompts may contain contextual information such as role-playing, fictional scenarios, research purposes, safety claims, defensive motivations, or additional task constraints. Such context can either constitute a legitimate part of the user's request or serve as contextual packaging that obscures the underlying objective. Therefore, the Rewrite Module does not assume that a particular type of context should always be retained or removed. Instead, it seeks to produce a more direct representation of the requested objective while preserving the semantic intent of the original prompt. This design allows potentially harmful intent hidden behind contextual packaging to become more explicit, while avoiding the assumption that sensitive terminology or contextual constraints necessarily imply malicious intent. 

The module further prohibits introducing new objectives, actions, entities, or intentions that are absent from the original prompt. Conversely, explicitly stated harmful objectives should not be weakened or transformed into benign objectives merely to reduce the perceived risk. The resulting rewritten variants therefore remain semantically grounded in the original prompt while providing alternative representations of the same underlying request. This distinction is important for subsequent risk comparison: the purpose of rewriting is not to make a prompt inherently safer or more harmful, but to reveal how its perceived risk changes when the underlying request is expressed more directly. 

Generating multiple rewritten variants provides several alternative semantic representations of the same original prompt. These variants reduce the dependence of subsequent risk assessment on a single paraphrase, while allowing the Scoring Module to identify whether the underlying intent consistently exhibits elevated risk across different representations. The resulting original prompt and five rewritten variants are subsequently passed to the Scoring Module for joint risk assessment.

\subsection{Scoring Module}

The Scoring Module evaluates the potential risk of the original prompt and its rewritten variants using multiple independent LLM-based safety scorers. Given the original prompt $x$ and its five rewritten variants ${x_1,x_2,\ldots,x_5}$, each prompt is independently evaluated by four scoring models. Each scorer assigns a continuous risk score in the range $[0,10]$, where a higher score indicates a higher level of potential harm.

For each prompt, the four scores are aggregated to obtain a joint risk score. Specifically, the joint risk score of the original prompt is computed as \begin{equation} S_0 = \frac{1}{4}\sum_{j=1}^{4}s_{0,j}, \end{equation} where $s_{0,j}$ denotes the score assigned to the original prompt by the $j$-th scorer. Similarly, the joint risk score of each rewritten variant $x_i$ is calculated as \begin{equation} S_i = \frac{1}{4}\sum_{j=1}^{4}s_{i,j}, \quad i\in{1,\ldots,5}. \end{equation}

Using multiple independent scorers provides a consensus-based assessment and reduces the influence of errors or biases from any individual evaluator. Since different LLM-based safety scorers may exhibit different sensitivities to harmful content and contextual formulations, relying on a single evaluator can lead to unstable risk judgments. Aggregating their assessments provides a more robust estimate of the perceived risk associated with each prompt representation.

In addition to the risk score of each individual rewritten variant, \tool computes the overall risk of the rewritten variants as \begin{equation} S_{\mathrm{var}} = \frac{1}{5}\sum_{i=1}^{5}S_i. \end{equation} The Scoring Module therefore produces both the original risk score $S_0$, the individual variant scores ${S_1,\ldots,S_5}$, and the overall rewritten-variant risk $S_{\mathrm{var}}$. These scores provide the basis for the Decision Module, which jointly considers absolute risk levels and the relative risk relationship between the original prompt and its rewritten variants to determine the appropriate defensive action.

\subsection{Decision Module}

The Decision Module converts the risk scores produced by the Scoring Module into defensive actions. Unlike a conventional threshold-based filter that relies solely on the absolute risk of the original prompt, \tool jointly considers the absolute risk levels and the relative risk relationship between the original prompt and its rewritten variants. This design enables \tool to distinguish between high-risk requests that should be intercepted, low-risk requests that can be passed through directly, and gray-zone requests that require more careful handling.

Let $S_0$ denote the joint risk score of the original prompt and $S_i$ denote the score of the $i$-th rewritten variant. Based on the scoring criteria, we use a low-risk threshold of 3.0 and a danger threshold of 6.0. The decision process first applies the danger threshold to identify high-risk requests. If $S_0 \geq 6.0$ or any rewritten variant satisfies $S_i \geq 6.0$, \tool directly intercepts the prompt. This hard-interception rule prevents requests with clearly elevated risk from being passed to the target model, regardless of their contextual formulation.

For prompts that do not trigger hard interception, \tool then identifies cases in which both the original prompt and all rewritten variants remain below the low-risk threshold. Specifically, if $S_0 < 3.0$ and $\max_i S_i < 3.0$, the original prompt is returned directly without modification. This case corresponds to requests whose alternative semantic representations consistently exhibit low perceived risk.

For the remaining gray-zone (3.0--5.9) cases, \tool compares the overall risk of the rewritten variants with that of the original prompt. As shown in Figure~\ref{fig:case}, we posit that, for a given malicious prompt wrapped in contextual packaging, its rewritten variants with the packaging removed will typically receive higher risk scores than the original prompt under the same scoring criteria. We define the \textit{risk factor} $\Delta$ as the risk difference between the rewritten variants and the original prompt: \begin{equation} \Delta=S_{\mathrm{var}}-S_0. \end{equation}

A positive $\Delta$ indicates that the rewritten representations exhibit higher perceived risk than the original formulation, suggesting that the contextual structure of the original prompt may be obscuring potentially harmful intent. When $\Delta>0$, \tool returns the highest-risk rewritten variant together with an explicit warning. Experiments show that when the highest-risk variant is harmful, it is more likely to be recognized and rejected by the target model. Therefore, \tool returns the highest-risk variant rather than the intuitively safer lowest-risk variant.

Conversely, when $\Delta\leq0$, the rewriting process does not reveal an increased risk trend. In this case, \tool preserves the original prompt but adds an explicit safety warning. This warning-based intervention allows the target model to retain the original request while providing additional safety guidance for cases whose risk remains ambiguous.

The resulting decision process applies different levels of intervention according to both absolute and relative risk. High-risk prompts are intercepted, consistently low-risk prompts are passed through unchanged, and gray-zone prompts are handled according to the risk factor revealed by semantic rewriting. A positive $\Delta$ routes the request toward the highest-risk rewritten representation with a warning, whereas a non-positive $\Delta$ preserves the original prompt with a warning. Importantly, gray-zone requests are not automatically rejected. Instead, \tool uses risk-aware prompt routing and explicit warnings to retain the legitimate functionality of potentially benign requests while providing the target model with additional safety guidance.

\section{Evaluation}
Our evaluation aims to answer the following research questions:

\noindent\emph{- RQ1 (Defense Effectiveness):} How effectively does \tool defend against direct harmful requests and black-box jailbreak attacks?

\noindent\emph{- RQ2 (Over-Refusal):} How effectively does \tool preserve benign requests while defending against harmful inputs?

\noindent\emph{- RQ3 (Ablation):} How does each component contribute to the overall effectiveness of \tool?

\noindent\emph{- RQ4 (Sensitivity):} How sensitive is \tool to the choice of the number of scoring models, and the number of rewritten variants?
\subsection{Settings}

\subsubsection{Evaluation Datasets \& Attacks}

We evaluate \tool from two complementary perspectives: defense effectiveness against harmful requests and resistance to over-refusal on benign but potentially sensitive requests. For harmful requests, we use \textbf{AdvBench}~\cite{zou2023universal}, a widely adopted benchmark of harmful behaviors and malicious prompts. To evaluate robustness against adversarial prompt manipulation, we further transform these prompts using three black-box jailbreak methods: \textbf{DeepInception}~\cite{li2023deepinception}, \textbf{CIPHER}~\cite{yuan2023gpt}, and \textbf{UNIATTACK}~\cite{wang2026automated}. These attacks conceal harmful intent through different prompt-level transformations, allowing us to evaluate defense against both explicit and adversarially contextualized harmful requests.

For over-refusal evaluation, we use \textbf{OR-Bench-Hard}~\cite{cui2024or}, a challenging subset containing benign requests that may appear harmful or sensitive on the surface. It evaluates whether \tool can distinguish genuinely harmful intent from legitimate requests involving sensitive topics. We randomly sample 300 prompts from AdvBench and 300 from OR-Bench-Hard. Each prompt is evaluated independently under the same defense and target-model configuration. For adversarial attacks, each AdvBench prompt is instantiated by each attack method and evaluated as an individual attack instance.

\subsubsection{Target Backbone Models}

To evaluate \tool across different levels of intrinsic safety alignment, we consider two target LLMs with substantially different safety behaviors: \textbf{Llama-3-8B-Uncensored}~\cite{hartford2024dolphin} and \textbf{DeepSeek-V4-Flash}. Llama-3-8B-Uncensored provides a challenging setting with limited intrinsic protection against harmful requests, allowing us to isolate the effectiveness of the external defense. In contrast, DeepSeek-V4-Flash is an instruction-following model with intrinsic safety alignment and refusal behaviors, enabling us to examine whether \tool remains effective while complementing existing model-level safety mechanisms.

We treat both target models as black-box systems and do not modify their parameters or internal representations. For each evaluation instance, the target model receives only the prompt produced by the corresponding defense method, ensuring that all methods are compared under the same target-model configuration. These two complementary backbones therefore allow us to assess whether \tool can independently mitigate harmful behavior on an uncensored model while providing additional protection without unnecessarily interfering with the safety mechanisms of an aligned model.

\subsubsection{Rewrite \& Score Models}

\tool separates semantic rewriting from risk assessment and requires no parameter updates to either component; both are implemented through system-level prompting with one-shot demonstrations. For semantic rewriting, we use \textbf{deepseek-v4-flash-thinking}~\cite{xu2026deepseek} as the rewriting model. For risk assessment, \tool employs four independent lightweight open-source LLMs as scoring models: \textbf{Gemma2-9B}~\cite{team2024gemma}, \textbf{Llama3-8B}~\cite{grattafiori2024llama}, \textbf{Mistral-Nemo}~\cite{mistral2024nemo}, and \textbf{Qwen2.5-7B}~\cite{hui2024qwen2}. Each scorer assigns a continuous risk score from 0 to 10, with higher scores indicating greater potential for harmful, illegal, or dangerous behavior. Each original prompt and rewritten variant is independently scored by all four models, and their scores are averaged to obtain the joint risk score.

\subsubsection{Baseline Defenses}

We compare \tool against three existing prompt-based jailbreak defenses: \textbf{Defensive Prompt Patch (DPP)~\cite{xiong2025defensive}}, \textbf{Intention Analysis (IA)~\cite{zhang2025intention}}, and \textbf{Self-Reminders (SR)~\cite{xie2023defending}}.

\textbf{DPP} introduces a defensive prompt patch designed to improve the robustness of LLMs against jailbreak attacks.

\textbf{IA} analyzes the intention underlying user inputs to identify potentially malicious requests before they are processed by the target model.

\textbf{SR} strengthens the target model's safety behavior by incorporating explicit self-reminder instructions into the prompt.

\subsubsection{Metrics}

We use the following metrics to evaluate \tool and the baselines.

\textbf{Defense Success Rate (DSR)}: The proportion of malicious attack prompts that are successfully prevented from eliciting harmful outputs from the target model.

\textbf{Over-Refusal Rate (ORR)}: The proportion of benign requests that are incorrectly rejected by the defense framework.

\begin{figure*}[t] 
    \centering
    \includegraphics[width=1\textwidth]{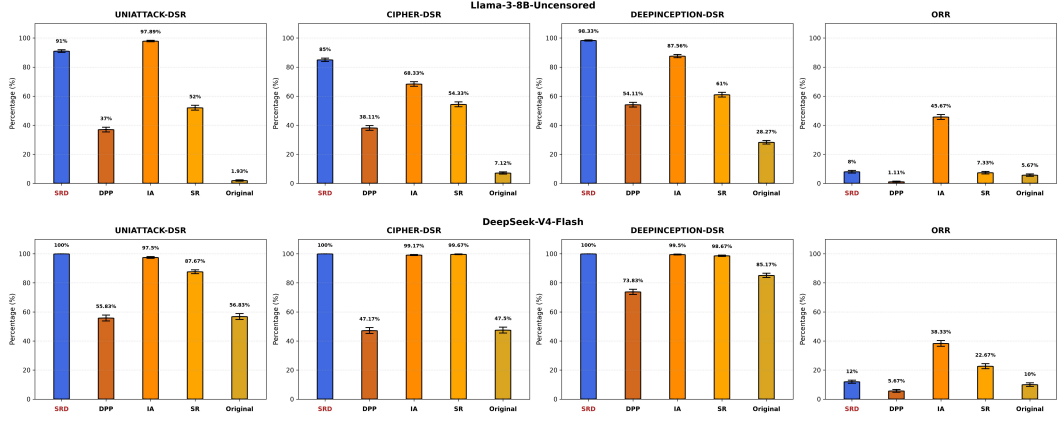}
    \caption{DSR and ORR results of \tool and three baselines on Llama-3-8B-Uncensored and DeepSeek-V4-Flash}
    \label{fig:result}
\end{figure*}

\subsection{Answer to RQ1 (Defense Effectiveness)}
Overall, as shown in Figure~\ref{fig:result}, \tool achieves consistently high Defense Success Rates (DSR) across UNIATTACK, CIPHER, and DeepInception, while maintaining substantially stronger performance than most baselines on the Llama-3-8B-Uncensored target. And on DeepSeek-V4-Flash, \tool is the only one that achieves a DSR of 100\% across all three attack scenarios, demonstrating that its defense remains highly effective against different attack formulations. These results indicate that \tool can reliably identify and intervene on harmful requests even when their malicious intent is concealed through different prompt-level attack strategies.

On Llama-3-8B-Uncensored, \tool achieves DSRs of 91.00\%, 85.00\%, and 98.33\% against UNIATTACK, CIPHER, and DeepInception, respectively. These results consistently outperform DPP and SR, while remaining competitive with IA, which achieves 97.89\%, 68.33\%, and 87.56\% on the three attacks. Notably, \tool performs particularly strongly against CIPHER and DeepInception, where its DSR exceeds the strongest baseline by 16.67 and 10.77 percentage points, respectively. On DeepSeek-V4-Flash, \tool achieves 100\% DSR for all three attacks, outperforming or matching all baselines in UNIATTACK and CIPHER and substantially outperforming them in DeepInception. \tool itself can identify nearly 93\% of malicious attacks, but does not choose to intercept all of them, primarily to reduce ORR. Specifically, \tool only intercepts inputs with a risk score of 6 or higher, while for inputs in the gray zone (3--5.9), it returns the prompt with an explicit safety warning. Since safety-aligned LLMs retain a strong degree of intrinsic capability to defend against malicious attacks, this design explains why \tool can achieve a low ORR while still attaining a 100\% DSR on safety-aligned models. The strong performance of \tool can be attributed to its multi-stage defense strategy: semantic rewriting exposes potentially concealed harmful intent, joint scoring aggregates judgments from multiple independent safety scorers, and the decision module combines absolute risk with the relative risk trend between the original and rewritten prompts. This allows \tool to detect harmful requests that may appear less risky in their original contextualized form rather than relying solely on surface-level characteristics.

For the original model, on Llama-3-8B-Uncensored, \tool increases ORR by only 2.33\% while improving DSR by 70.06\%-89.07\% across the three attacks. On DeepSeek-V4-Flash, \tool incurs only a 2.00\% increase in ORR while improving DSR by 14.83-52.50\%.

\subsection{Answer to RQ2 (Over-Refusal)}
Figure~\ref{fig:result} also compares the Over-Refusal Rate (ORR) of \tool and the baseline defenses on OR-Bench-Hard. Overall, \tool achieves a substantially lower ORR than the intention-based baseline IA while maintaining much higher DSR against harmful inputs. On Llama-3-8B-Uncensored, \tool obtains an ORR of only 8.00\%, compared with 45.67\% for IA. On DeepSeek-V4-Flash, \tool achieves an ORR of 12.00\%, again substantially lower than IA's 38.33\%. These results demonstrate that \tool does not simply improve defense effectiveness by conservatively rejecting a large fraction of user requests. Instead, it preserves a considerably larger portion of benign inputs while still providing strong protection against harmful requests.

The joint DSR--ORR results further illustrate this advantage. On Llama-3, \tool achieves DSRs of 91.00\%, 85.00\%, and 98.33\% against UNIATTACK, CIPHER, and DeepInception, respectively, while maintaining an ORR of only 8.00\%. In contrast, IA achieves a higher DSR on UNIATTACK (97.89\%) but incurs an ORR of 45.67\%, indicating that its stronger defense on this attack comes with a substantial loss in benign-request preservation. On DeepSeek-V4-Flash, \tool achieves 100\% DSR across all three attack scenarios with an ORR of 12.00\%, whereas IA obtains 97.50\%, 99.17\%, and 99.50\% DSR with an ORR of 38.33\%. This consistent reduction in ORR suggests that \tool's rewriting and relative-risk-based decision mechanism provides a more selective form of intervention. Instead of treating potentsially sensitive inputs as inherently harmful, \tool examines whether semantic rewriting reveals a meaningful increase in perceived risk and only routes gray-zone requests more aggressively when the risk trend provides sufficient evidence.

On Llama-3-8B-Uncensored, \tool improves the average DSR by 30.30\% over the baselines while reducing ORR by 10.04\%. On DeepSeek-V4-Flash, \tool achieves a further 15.67\% improvement in average DSR while reducing ORR by 10.22\%.

\subsection{Answer to RQ3 (Ablation)}

\begin{table*}
\centering
\caption{Ablation study results on Llama-3-8B-Uncensored and DeepSeek-V4-Flash. (The top-performing results are highlighted in \textbf{Bold}.)}
\label{tab:ablation}

\resizebox{\textwidth}{!}{
\begin{tabular}{|l|c|c|c|c|}
\hline

\textbf{Method}
& \textbf{UNIATTACK-DSR}
& \textbf{DEEPINCEPTION-DSR}
& \textbf{CIPHER-DSR}
& \textbf{ORR} \\
\hline

\multicolumn{5}{|c|}{\cellcolor[HTML]{EFEFEF} \textbf{Llama-3-8B-Uncensored}} \\
\hline

Full-System
& $91 \pm 0.95\%$
& $\textbf{98.33} \pm 0.43\%$
& $\textbf{85} \pm 1.19\%$
& $\textbf{8} \pm 0.91\%$ \\
\hline

w/o Rewrite Module
& $41.58 \pm 3.46\% \downarrow$
& $47.03 \pm 3.51\% \downarrow$
& $63.37 \pm 2.77\% \downarrow$
& $8.47 \pm 0.99\% \uparrow$ \\
\hline

w/o Joint Scoring
& $\textbf{100}\% \uparrow$
& $96 \pm 1.39\% \downarrow$
& $75 \pm 3.06\% \downarrow$
& $37 \pm 3.42\% \uparrow$ \\
\hline

w/o Decision Module
& $71 \pm 3.21\% \downarrow$
& $93 \pm 1.8\% \downarrow$
& $74 \pm 3.1\% \downarrow$
& $24.5 \pm 3.04\% \uparrow$ \\
\hline

\multicolumn{5}{|c|}{\cellcolor[HTML]{EFEFEF} \textbf{DeepSeek-V4-Flash}} \\
\hline

Full-System
& $\textbf{100}\%$
& $\textbf{100}\%$
& $\textbf{100}\%$
& $\textbf{12} \pm 1.09\%$ \\
\hline

w/o Rewrite Module
& $49.01 \pm 3.51\% \downarrow$
& $90.59 \pm 2.05\% \downarrow$
& $53.96 \pm 3.51\% \downarrow$
& $16.5 \pm 2.62\% \uparrow$ \\
\hline

w/o Joint Scoring
& $100\%$
& $100\%$
& $89 \pm 2.21\% \downarrow$
& $50 \pm 3.54\% \uparrow$ \\
\hline

w/o Decision Algorithm
& $71.5 \pm 3.19\% \downarrow$
& $100\%$
& $67 \pm 3.32\% \downarrow$
& $32.5 \pm 3.31\% \uparrow$ \\
\hline

\end{tabular}
}
\end{table*}
Table~\ref{tab:ablation} presents the ablation results on Llama-3-8B-Uncensored and DeepSeek-V4-Flash. Overall, the complete \tool achieves the highest average DSR while maintaining the lowest ORR across both target models, demonstrating that the three components contribute complementarily to the overall DSR--ORR trade-off. Removing any component leads to a noticeable degradation in either defense effectiveness, over-refusal control, or both.

\textbf{w/o Rewrite Module.} Removing the Rewrite Module causes the most substantial degradation in DSR. On Llama-3-8B-Uncensored, the average DSR drops from 91.44\% to 50.66\%, while on DeepSeek-V4-Flash it decreases from 100\% to 64.52\%. The degradation is particularly pronounced for UNIATTACK and DeepInception, whose contextual constructions can effectively conceal the underlying harmful intent from both the scoring models and the target model. Without semantic rewriting, the defense relies primarily on the original contextualized representation, making it substantially more difficult to expose the malicious objective. Interestingly, w/o Rewrite Module maintains the second-lowest ORR, at 8.47\% and 16.50\% on the two target models, respectively. This result indicates that the low ORR is achieved partly by adopting a less aggressive intervention strategy, but at the cost of substantially reduced defense effectiveness. Thus, the Rewrite Module is critical for exposing concealed malicious intent rather than simply increasing the overall intervention rate.

\textbf{w/o Joint Scoring.} In this ablation, the four-model scoring ensemble is replaced by a single \textbf{Llama3-8B} scorer. This variant exhibits an interesting anomaly on Llama-3-8B-Uncensored: its UNIATTACK DSR reaches 100\%, exceeding the 91\% achieved by the full system. This does not indicate that single-model scoring is intrinsically more effective. Under our scoring rubric, Llama3-8B exhibits an unusually strong sensitivity to potentially harmful inputs and tends to assign elevated risk scores, resulting in substantially more aggressive interception. This behavior is reflected in its ORR, which increases sharply to 37\% on Llama-3-8B-Uncensored and 50\% on DeepSeek-V4-Flash. Therefore, the apparent DSR improvement is accompanied by a substantial loss in benign-request preservation.

This result is also related to the interception behavior of \tool. Under the UNIATTACK templates, \tool intercepts 100\% of attack instances before they reach the target model, while the corresponding interception rate for DeepInception is 86\%. Consequently, all UNIATTACK instances and most DeepInception instances are directly rejected rather than passed to the target model. This means that the remaining attack instances actually reaching the target model constitute only a small subset, which helps explain why the w/o Joint Scoring variant can still achieve very high DSR, including 100\% on some settings. The result therefore highlights the importance of multi-model scoring for controlling the aggressiveness of risk estimation rather than simply maximizing interception.

\textbf{w/o Decision Algorithm.} Removing the Decision Module substantially reduces DSR while increasing ORR on both target models. On Llama-3-8B-Uncensored, the average DSR decreases to 79.33\% and ORR increases to 24.50\%; on DeepSeek-V4-Flash, the corresponding values are 79.50\% and 32.50\%. Although the joint scoring mechanism remains available, the defense loses the risk-adaptive routing mechanism that distinguishes consistently low-risk inputs from ambiguous gray-zone cases. In particular, without the Decision Module, sensitive inputs that would otherwise be handled selectively according to the relative risk between the original and rewritten prompts are subjected to less discriminative intervention, resulting in substantially higher over-refusal. This demonstrates that accurate risk estimation alone is insufficient: the decision mechanism is necessary to translate risk signals into appropriate actions and maintain a favorable DSR--ORR trade-off.

Overall, the ablation results confirm the complementary roles of the three components. The Rewrite Module is essential for exposing harmful intent hidden by contextual packaging, Joint Scoring provides more stable and less model-dependent risk estimation, and the Decision Module converts these signals into risk-adaptive interventions that control over-refusal. The full system therefore achieves the strongest overall balance between defense effectiveness and benign-request preservation.

\subsection{Answer to RQ4 (Sensitivity)}
\begin{wraptable}{r}{0.48\columnwidth}
\centering
\caption{Sensitivity analysis on Llama-3-8B-Uncensored under CIPHER.}
\label{tab:sensitivity}
\small
\begin{tabular}{cccc}
\hline
\textbf{Setting} & \textbf{Number} & \textbf{DSR} & \textbf{ORR} \\
\midrule \multirow{4}{*}{Rewrite Variants}
& 1 & 80\% & 24\% \\
& 3 & 81\% & 16\% \\
& \textbf{5} & \textbf{85\%} & \textbf{8\%} \\
& 7 & 82\% & 12\% \\
\midrule \multirow{4}{*}{Scoring Models}
& 1 & 75\% & 37\% \\
& 2 & 71\% & 17\% \\
& 3 & 74\% & 26\% \\
& \textbf{4} & \textbf{85}\% & \textbf{8}\% \\
& 5 & 66\% & 32\% \\
\hline
\end{tabular}
\end{wraptable}

Table~\ref{tab:sensitivity} shows that for the number of rewritten variants, increasing the number from 1 to 5 gradually improves the DSR--ORR trade-off, with five variants achieving the highest DSR (85\%) and the lowest ORR (8\%). Increasing the number further to seven does not provide additional benefit, reducing DSR to 82\% and increasing ORR to 12\%. This suggests that five rewritten variants provide a favorable balance between representation diversity and decision stability. For the number of scoring models, four scorers achieve the strongest overall performance, with 85\% DSR and 8\% ORR. Using fewer scorers results in either lower DSR or higher ORR, while adding a fifth scorer, \textbf{Phi-3-Mini}~\cite{haider2024phi}, substantially degrades performance to 66\% DSR and 32\% ORR. This result indicates that increasing the ensemble size does not necessarily improve risk estimation, as additional scorers may provide less compatible risk judgments. Overall, the sensitivity analysis supports the use of five rewritten variants and four scoring models as an effective configuration for \tool.

Furthermore, \tool supports flexible model updates within its defense framework. As more capable models for risk scoring or semantic rewriting become available, the performance of \tool could potentially be further improved.

\section{Discussion}
\textbf{Deployment Overhead.}
\tool introduces additional computational overhead because its scoring module relies on four independent local LLM-based evaluators rather than a lightweight binary classifier. These evaluators must process both the original prompt and its rewritten variants, and therefore require additional memory and computational resources during deployment. Moreover, the scoring models cannot be arbitrarily replaced by extremely lightweight classifiers that only output binary labels, since \tool relies on continuous risk scores and requires the evaluators to follow a predefined scoring rubric. Consequently, deploying \tool may require substantially more hardware resources than conventional input-level defenses based on a single classifier or rule-based filter. Exploring more efficient scoring architectures, such as smaller instruction-following models or hybrid lightweight evaluators, is an important direction for future work.

\textbf{Ambiguity of Safety Boundaries.}
A second limitation arises from the inherent ambiguity between harmful requests and benign but sensitive or dual-use requests. Datasets such as OR-Bench contain many prompts that deliberately combine potentially sensitive behaviors with legitimate contextual goals, making it difficult to establish a sharp boundary between harmful intent and acceptable use under a unified risk criterion. Although \tool mitigates this ambiguity by jointly considering the absolute risk and the relative risk between the original prompt and its rewritten variants, the distinction cannot always be perfectly resolved. As a result, \tool cannot simultaneously achieve zero over-refusal and perfect defense success: increasing sensitivity to hidden harmful intent may inevitably affect some borderline benign requests, while relaxing the defense may leave some harmful requests insufficiently detected. This limitation reflects a broader challenge in LLM safety evaluation, where harmfulness and benign utility often form a continuous spectrum rather than a strictly separable binary boundary.

\section{Related Work}
Input-level defense methods detect or transform suspicious prompts before generation. SmoothLLM~\cite{robey2023smoothllm} applies randomized perturbations to multiple input copies and aggregates their predictions, while other methods use safety prompts or self-reminders to reinforce safety policies~\cite{xie2023defending,xiong2025defensive}. Although effective for improving DSR, these approaches primarily rely on surface-level cues and may increase ORR by treating suspicious inputs conservatively. In contrast, \tool exposes potentially hidden intent through semantic rewriting and applies fine-grained risk assessment to better preserve benign or dual-use requests.

Intention-based defenses explicitly analyze the essential intent of a request before generation~\cite{li2025unraveling}. Intention Analysis (IA)~\cite{zhang2025intention}, for example, identifies the query's essential intention and generates a response conditioned on this analysis. However, IA relies on a single inferred intention and does not compare the original prompt with multiple semantic representations. \tool instead generates multiple rewrites and jointly evaluates them with the original prompt, providing additional evidence for distinguishing concealed harmful intent from benign contextual variations.

LLM-based safety judges and multi-model defenses improve the reliability of safety assessment. AutoDefense~\cite{zeng2024autodefense}, for example, employs multiple LLM agents to collaboratively analyze potentially harmful inputs. However, existing approaches generally assess a single prompt representation using absolute safety judgments, without continuous risk scoring or explicit relative-risk comparison. \tool addresses these limitations by jointly scoring the original prompt and its rewrites with multiple independent scorers under a unified continuous risk rubric. This consensus-based assessment reduces dependence on individual judges, while relative-risk comparison provides an additional signal for adaptive defensive routing.


\section{Conclusion}
We presented \tool, a parameter-free, black-box defense framework for mitigating LLM jailbreak attacks while reducing unnecessary refusal of benign requests. The key idea is to treat semantic rewriting as an additional source of evidence for safety assessment: rather than judging a prompt solely from its original contextualized form, \tool generates multiple semantically faithful representations to expose the underlying user intent. These representations are jointly evaluated by multiple independent LLM-based safety scorers, and a decision module combines absolute risk with the relative risk change between the original and rewritten prompts to perform risk-adaptive routing.

Our experiments across UNIATTACK, CIPHER, and DeepInception demonstrate that \tool provides consistently strong defense effectiveness across target models with substantially different intrinsic safety behaviors. It achieves an average DSR of 91.44\% on Llama-3-8B-Uncensored and 100\% on DeepSeek-V4-Flash, while maintaining low ORRs of 8.00\% and 12.00\%, respectively. These results indicate that \tool can identify concealed harmful intent without relying on uniformly aggressive rejection. The ablation study further confirms the complementary roles of its components: rewriting is essential for exposing contextual camouflage, joint scoring improves robustness against individual evaluator bias and excessive sensitivity, and the decision module is necessary for selectively handling gray-zone inputs.

Overall, \tool demonstrates that effective jailbreak defense does not require modifying the target model or applying a uniformly restrictive safety boundary. By combining semantic intent exposure, consensus-based risk assessment, and relative-risk-aware decision making, \tool achieves a more favorable balance between defense effectiveness and benign-request preservation. This provides a practical direction for deploying adaptive safety layers around black-box LLMs while retaining their general-purpose utility.

\bibliography{citation}
\bibliographystyle{iclr2027_conference}

\end{document}